\documentclass[prb,twocolumn,showpacs]{revtex4}
\usepackage{graphicx}
\usepackage{hyperref}

\begin{document}

\title{Unification Theory of Angular Magnetoresistance Oscillations in Quasi-One-Dimensional Conductors}
\author{Si Wu}
\author{A.G. Lebed$^*$}
\affiliation{Department of Physics, University of Arizona, 1118 E.
4th St., Tucson, AZ 85721}

\date{\today}

\begin{abstract}

We present a unification theory of angular magnetoresistance oscillations,
experimentally observed in quasi-one-dimensional organic conductors, by 
solving the Boltzmann kinetic equation in the extended Brillouin zone. 
We find that, at commensurate directions of a magnetic field, resistivity 
exhibits strong minima. 
In two limiting cases, our general solution reduces to the results, previously 
obtained for the  Lebed Magic Angles and Lee-Naughton-Lebed oscillations. 
We demonstrate that our theoretical results are in good qualitative and quantitative
agreement with the existing measurements of resistivity in (TMTSF)$_2$ClO$_4$
conductor.
\end{abstract}


\maketitle

\section{Introduction}

Magnetic properties of quasi-one-dimensional (Q1D) organic conductors
have been intensively studied both experimentally and theoretically since 
a discovery of the so-called Field-Induced Spin-Density-Wave (FISDW) 
phase diagrams$^{1,2}$.
For open electron orbits, where the Landau levels quantization is impossible, as
theoretically shown$^{1,3-5}$, 
other quantum effects - the Bragg reflections
of electrons from the Brillouin zones - play an important role.
In the simplest situation, where magnetic field is perpendicular to conducting 
layers of Q1D conductors, the Bragg reflections are 
demonstrated$^{3-5}$ to be 
responsible for the FISDW phases formation.
In an inclined magnetic field, perpendicular to conducting chains, a more 
complicated interference picture of electrons, moving in the extended Brillouin 
zones, appears. 
As shown in Refs. 6 and 7, it results in the constructive interference of electron waves
in some many-body effects for some special commensurate directions of a magnetic field, 
which are called the Lebed Magic 
Angles (LMA).

The LMA effects were experimentally discovered in Refs. 8-13 and are 
observed in a number of organic conductors$^{14-29}$, 
which possess Q1D parts 
in their Fermi surfaces (FS).
Note that, instead of maxima of resistivity due to electron-electron scattering, 
predicted in Ref. 7, the experiments$^{8-29}$ demonstrate clear minima at the 
LMA directions of a magnetic field.
In important theoretical contributions$^{30,31}$, it was shown that constructive 
interference effects$^{6,7}$ could appear in such one-body phenomenon as a
residual resistivity due to impurities.
As a result, minima of resistivity component, perpendicular to conducting layers, 
were theoretically found at the LMA directions of a magnetic field$^{30,31}$.
Nevertheless, theoretical model$^{30}$ predicted weak (i.e., exponentially small) 
magnitudes of the LMA minima, whereas Ref. 31 was based on an incorrect 
solution of the
Boltzmann kinetic equation.

Recently, a correct solution of the Boltzmann equation for a magnetic field,
perpendicular to conducting chains of a Q1D conductor, was found and
the existence of the strong LMA minima in perpendicular to conducting layers 
component of resistivity was firmly 
established$^{32}$.
In addition, a quantum mechanical variant of the theory was 
suggested$^{33}$.
Theory$^{33}$ reveals quantum interference nature of periodic solutions of
the Boltzmann kinetic equation in the extended Brillouin zone$^{32}$, which lead
to the appearance of the LMA effects.
According to Ref. 33, the LMA minima of resistivity appear due to changes 
of electron wave functions dimensionality from 1D into 2D at commensurate
directions of a magnetic field$^{6,7}$ due to constructive interference 
effects.
These interference effects appear between velocity component, perpendicular
to conducting layers, and the density of states$^{33}$.
Note that the theory$^{32,33}$, which is a limiting case of the unification theory,
suggested in this paper, can explain the experimental observations of the LMA
effects only in resistivity$^{8-29}$.
As to the observations of anomalously strong LMA phenomena in the Nernst$^{34-38}$ 
and Hall$^{39}$ effects, their explanations may need a different theoretical
approach.

As experimentally discovered$^{40-43}$, the LMA-like magnetoresistance 
minima with enhanced magnitudes are observed$^{40-46}$ in experimental 
geometry, where magnetic field is inclined with respect to conducting
chains of a Q1D conductor.
At first, these angular oscillations were interpreted$^{42,43}$ in terms of the
LMA phenomenon.
Later, it became clear$^{47-50}$ that, although they are related to the
LMA effects, their concrete physical meaning is quite different.
Below, we call the above mentioned angular oscillations of resistivity the
Lee-Naughton-Lebed (LNL) ones.

Theory of the LNL phenomenon, based on a solution of the Boltzmann
kinetic equation in the extended Brillouin zone, was suggested in 
Refs. 47, 48.
In Refs. 47, 49, theory of the LNL oscillations was extended to the 
so-called weak non Fermi-liquid quantum case$^{47}$.
In Refs. 44, 45, 50, the quantum theory of the LNL oscillations was suggested
for the Fermi-liquid case and their quantum interference nature was
revealed.
According to Refs. 44, 45, 50, the LNL angular oscillations are due to changes
of electron wave functions dimensionality from 1D into 2D at some
commensurate directions of a magnetic field due to constructive
interference effects.
In contrast to case of the LMA oscillations, these interference effects
appear between two velocity components, perpendicular to the conducting
chains.
By present time, the physical origin of the LNL oscillations has been firmly
established$^{44-50}$ and a comparison of the theoretical results with the
existing experimental data have been made$^{44,45,50}$.
These include theories of the LNL phenomenon in the presence of anion
ordering potentials$^{44,45,51}$ and theory$^{52}$, connecting the LNL oscillations
with the Aharonov-Bohm effect.

The goal of our paper is to suggest an analytical unified theory, which 
describes both the LMA and LNL phenomena as its limiting cases.
Note that the suggested theory also describes the so-called 
Danner-Kang-Chaikin (DKC) oscillations$^{53}$ and Third Angular Effect 
(TAE)$^{54-56}$.
To derive an analytical expression for resistivity component, perpendicular
to conducting layers, we analytically solve the Boltzmann kinetic equation 
in the extended Brillouin zone in a magnetic field, inclined with respect to
conducting chains of a Q1D conductor.
We find strong minima of resistivity, corresponding to the appearance
of periodic solutions of the Boltzmann equation at the LMA and LNL
commensurate directions of a magnetic field.
A comparison of our theoretical results with the experimental data on
(TMTSF)$_2$ClO$_4$ conductor$^{57}$ shows good qualitative and quantitative
agreement (see Figs. \ref{LNLFig}, \ref{MAFig}).

\section{Unification Theory}

Let us consider a Q1D conductor with the following electron spectrum,
\begin{eqnarray}
\label{Spectrum} &&\varepsilon^{\pm}({\bf p})=\pm v_x(p_yb^*)[p_x\mp
p_x(p_y)] - 2t_c\cos(p_zc^*),\nonumber \\ 
&&p_x(p_y)=p_F-2t_b g(p_yb^*)/v_F.
\end{eqnarray}
[Here, $+(-)$ stands for right (left) sheet of the FS;
$v_F$ and $p_F$ are the Fermi velocity and Fermi momentum along the most
conducting ${\bf x}$ axis, respectively; $t_b$ and $t_c$ are
the hopping integrals along ${\bf y}$ and ${\bf z}$ 
axes.]
Note that in Eq. (\ref{Spectrum}), in accordance with Refs. 31-33, $p_y$ dependence of the 
velocity along conducting chains, $v_x(p_y)$, is taken into 
account. 
For most Q1D conductors, we can choose $g(p_yb^*)=\cos(p_yb^*)$ in Eq. (\ref{Spectrum}), 
but for an important exception of (TMTSF)$_2$ClO$_4$ conductor with an 
anion ordering, we have to use
$g(p_yb^*)=[\cos^2(p_yb^*)+(\Delta/2t_b)^2]^{1/2}$, where 
$\Delta$ is the the so-called anion 
gap$^{1,2,44,45}$.

Below, we show that, when a Q1D conductor (\ref{Spectrum}) is placed in a tilted magnetic field,
\begin{equation}
\label{H} {\bf
H}=H(\cos\theta\cos\varphi,\cos\theta\sin\varphi,\sin\theta),
\end{equation}
and a weak electric field perpendicular to the conducting 
$({\bf x},{\bf y})$ plane, then at certain orientations of the field, 
\begin{equation}
\label{commensurate}
\sin\varphi=n\left(\frac{b^*}{c^*}\right)\tan\theta,
\end{equation}
where $n$ is an integer, resistivity, $\rho_{zz}(H,\theta,\varphi)$, 
exhibits  strong minima. 

The Boltzmann
kinetic equation in crossed electric and magnetic fields in the so-called
$\tau$-approximation can be written in a standard way,
\begin{equation}
\label{BE1} \Big[e{\bf E}+ \Big( \frac{e}{c} \Big) [{\bf v}\times{\bf
H}] \Big]\frac{\partial f({\bf p})}{\partial{\bf p}}=-\frac{f({\bf
p})-f_0({\bf p})}{\tau} .
\end{equation}
[In Eq. (\ref{BE1}), $f({\bf p})$ is an electron distribution function and $\tau$ is
a relaxation time.] 
After the standard approximation,
\begin{equation}
f({\bf p})=f_0({\bf p})-\frac{\partial f_0({\bf p})}{\partial
\varepsilon}\Psi(p_y,p_z),
\end{equation}
the Boltzmann equation (\ref{BE1}) can be written as$^{58}$,
\begin{equation}
e{\bf E}\cdot{\bf v}- \Big( \frac{e}{c} \Big) [{\bf v}\times{\bf H}]
\cdot\frac{\partial\Psi(p_y,p_z)}{\partial{\bf p}}= \frac{\Psi(p_y,p_z)}{\tau} ,
\end{equation}
where  $f_0({\bf p})$ is the  Fermi-Dirac distribution 
function.
[Note that, in Eq.(6), we use independent variables $(\epsilon, p_y,p_z)$$^{58}$, instead
of $(p_x,p_y,p_z)$, where energy $\epsilon$ is conserved in the absence of 
electric field.
Since both electric field and temperature are supposed to be small, the electric
conductivity is defined by electrons, located in the near vicinity of the Q1D FS.
Therefore, the distribution function $\Psi(p_y,p_z)$ can be taken at $\epsilon = \epsilon_F$ and, thus, does not depend on energy in Eq.(6).] 
Taking into account that 
${\bf v}=\partial\varepsilon({\bf p})/\partial{\bf p}$, we can rewrite the Boltzmann 
equation as,
\begin{eqnarray}
\label{BE2} 
&&eEv_z^0\sin z - \omega_b(y,\theta)\frac{\partial
\Psi(y,z)}{\partial y} - \omega_c(y,\theta,\varphi)\frac{\partial
\Psi(y,z)}{\partial z} \nonumber\\
&&\ \ \ \ \ \ \  - \omega_c^*(\theta,\varphi) g^{\prime}(y)\frac{\partial
\Psi(y,z)}{\partial z}=\frac{\Psi(y,z)}{\tau}.
\end{eqnarray}
To simplify the notations, we define the following dimensionless parameters: 
$y=p_yb^*$, $z=p_zc^*$, and the following frequency variables: 
\begin{eqnarray}
\label{frequency}
&&\omega_b(y,\theta) = \Big( \frac{e}{c} \Big) v_x(y)Hb^*\sin\theta,\nonumber\\
&&\omega_c(y,\theta,\varphi) = \Big( \frac{e}{c} \Big) v_x(y)Hc^*\cos\theta\sin\varphi, \\
&&\omega_c^*(\theta,\varphi)=
\Big( \frac{e}{c} \Big) v_y^0Hc^*\cos\theta\cos\varphi,\nonumber
\end{eqnarray}
where $v_y^0=2t_bb^*$ and $v_z^0=2t_cc^*$; $\hbar = 1$.

It is important that the partial differential equation (\ref{BE2}) can be analytically solved
(see the Appendix A),
\begin{eqnarray}
\label{Solution2} 
&&\Psi(y,z)=eEv_z^0\int_{-\infty}^0\frac{{\rm
d}u}{\omega_b(y+u,\theta)}\nonumber\\
&&\cdot\sin
\Bigg[z-\int_u^0\frac{{\rm
d}v}{\omega_b(y+v,\theta)}\Big[\omega_c(y+v,\theta,\varphi)\nonumber\\
&&+\omega_c^*(\theta,\varphi) g^{\prime}
\left(y+v\right)\Big]\Bigg]\nonumber\\
&&\cdot\exp\left[-\int_u^0\frac{{\rm d}v}{\tau\omega_b(y+v,\theta)}\right].
\end{eqnarray}
Since the solution (\ref{Solution2}) of the Boltzmann equation (\ref{BE2}) is known, we can express
electric current density along ${\bf z}$ axis, perpendicular to conducting layers, 
in the following way,
\begin{equation}
\label{current}
j_z\sim\int{\rm d}y\int{\rm d}zv_z^0\sin z\frac{\partial f_0({\bf p})}
{\partial \varepsilon}\Psi(y,z).
\end{equation}
In Eq. (\ref{current}), we omit an exact factor since conductivity in a magnetic field
is scaled below to its value in the 
absence of the field. 
Notice that $\partial f_0({\bf p})/\partial\varepsilon$ in Eq. (\ref{current}) is the density of
states at the FS, which is proportional to $1/v_x(y)$. From Eqs. (\ref{Solution2}), (\ref{current}), 
we obtain the following result for inter-layer conductivity
after some calculations,
\begin{eqnarray}
\label{conductance}
&&\sigma_{zz}(H,\theta,\varphi)\sim\int\frac{{\rm
d}y}{v_x(y)}\int_{-\infty}^0\frac{{\rm d}u}{\omega_b(y+u)}\nonumber\\
&&\cdot\cos
\Bigg[\int_u^0\frac{{\rm
d}v}{\omega_b(y+v,\theta)}\Big[\omega_c(y+v,\theta,\varphi)\nonumber\\
&&+\omega_c^*(\theta,\varphi)
g^{\prime} \left(y+v\right)\Big]\Bigg]\nonumber\\
&&\cdot\exp\left[-\int_u^0\frac{{\rm
d}v}{\omega_b(y+v)\tau}\right].
\end{eqnarray}
[Note that, in a Q1D case resistivity component, perpendicular to conducting
layers, is $\rho_{zz} (H,\theta,\varphi) =1/\sigma_{zz} (H,\theta,\varphi)$.]
Eqs.(9)-(11) and comparison of Eq.(11) limiting cases with the experimental 
data, obtained on (TMTSF)$_2$ClO$_4$ conductor$^{57}$, are the main results of 
our paper.

\section{Limiting cases}

In this section, we consider two different experimental settings, 
which correspond to the LNL and LMA phenomena, 
repectively.

\subsection{LNL Oscillations}

In experiments, where the LNL oscillations are observed, magnetic field is 
tilted away from $({\bf y},{\bf z})$ plane. 
In the case, where $\theta \ll 90^0$, the $p_y$ dependence of the Fermi 
velocity is not significant, and, therefore, we can everywhere replace $v_x(p_y)$ 
into $v_F$. 
This significantly simplifies the formula for conductivity (\ref{conductance}), 
\begin{eqnarray}
\label{LNL1}
&&\sigma_{zz}(H,\theta,\varphi)\sim\int_{-\infty}^0{\rm d}z\exp(z)
\int_0^{2\pi}\frac{{\rm
d}y}{2\pi}\cos\Big(\omega_c(\theta,\varphi)\tau z\nonumber\\
&&\ \ \ \ +\frac{\omega_c^*(\theta,\varphi)}{\omega_b(\theta)}\left[g(y)-g
\left(y+\omega_b(\theta)\tau z\right)\right]\Big),
\end{eqnarray}
where the frequency variables are defined in Eq. (\ref{frequency})
with $v_x(y) = v_F$.
The existence of resistivity minima at the commensurate directions 
of a magnetic field (\ref{commensurate}) can be understood
from Eq. (\ref{LNL1}). 
Since $g(y)$ is a periodic function of $y$, at the commensurate directions 
[where a condition $\omega_c(\theta,\varphi)=n\omega_b(\theta)$ is satisfied], 
constructive interference effects increase the integral 
(\ref{LNL1}), which gives the minima in $\rho_{zz}(H,\theta,\varphi)$.

\begin{figure}[t]
\includegraphics[width=.4\textwidth]{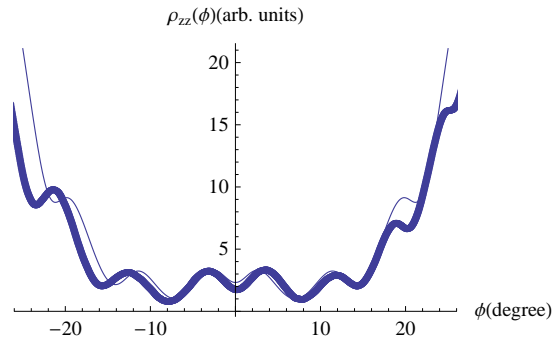}
\caption{\label{LNLFig}Comparison of the theory (\ref{LNL1}) and the experiment\cite{HaNaught} for
the Lee-Naughton-Lebed (LNL) oscillations in 
(TMTSF)$_2$ClO$_4$. 
Thin solid curve is calculated from Eq. (\ref{LNL1}) at 
$\theta=7^{\circ}$, with band parameters being 
$t_a/t_b=8.5$ and $\Delta/2t_b=0.1$. 
For calculations, we have used $\omega_b(0)\tau=15$ 
and $c^*=2b^*$. 
Numerical results are in good qualitative and quantitative agreement with 
the experimental data (thick solid line) in a broad range of angle $\phi$.}
\end{figure}

In Fig. \ref{LNLFig}, we show both the theoretical results, obtained by means of
numerical calculations of Eq. (\ref{LNL1}), and the experimental data$^{57}$,
obtained on (TMTSF)$_2$ClO$_4$ conductor.
Note that the minima of resistivity appear only for even values of n
in Eq. (\ref{commensurate}) due to the existence of the anion ordering gap, $\Delta$,
in (TMTSF)$_2$ClO$_4$ electron spectrum, 
$g(p_yb^*)=[\cos^2(p_yb^*)+(\Delta/2t_b)^2]^{1/2}$ 
[see Eq.(\ref{Spectrum})]. 

Most Q1D conductors [e.g., (TMTSF)$_2$PF$_6$ and 
$\kappa$-(ET)$_2$Cu(NCS)$_2$] can be described by the electronic 
spectrum (\ref{Spectrum}) with 
$g(y)=\cos y$. 
In this case, Eq. (\ref{LNL1}) reduces to, 
\begin{eqnarray}
\label{LNL2}
&&\sigma_{zz}(H,\theta,\varphi)\sim\int_{-\infty}^0{\rm d}z\exp(z)
\int_0^{2\pi}\frac{{\rm
d}y}{2\pi}\cos\Big[\omega_c(\theta,\varphi)\tau z\nonumber\\
&&\ \ \ \ +\frac{\omega_c^*(\theta,\varphi)}{\omega_b(\theta)}\left(\cos(y)-\cos
\left(y+\omega_b(\theta)\tau z\right)\right)\Big].
\end{eqnarray}
Using the identity,
\begin{equation}
\exp(iz\cos\theta)=\sum_{n=-\infty}^{+\infty}J_n(z)i^n{\rm e}^{in\theta},
\end{equation}
we can further simplify Eq. (\ref{LNL2}),
\begin{eqnarray}
\label{LNL3}
&&\sigma_{zz}(H,\theta,\varphi)\sim\int_{-\infty}^0{\rm d}z\exp(z)\\
&&\ \ \ \cdot\sum_{n=-\infty}^{+\infty}J_n^2\left(\frac{\omega_c^*(\theta,\varphi)}{\omega_b(\theta)}\right)
\cos\left((\omega_c(\theta,\varphi)-n\omega_b(\theta))\tau z\right)\nonumber.
\end{eqnarray}
Finally, Eq. (\ref{LNL3}) can be transformed into a simple analytical form,
\begin{equation}
\label{LNL4}
\frac{\sigma_{zz}(H,\theta,\varphi)}{\sigma_{zz}(H=0)}=\sum_{n=-\infty}^{\infty}
\frac{J_n^2\left[\omega_c^*(\theta,\varphi)/\omega_b(\theta)\right]}{1+\tau^2\left[\omega_c(\theta,\varphi)-n\omega_b(\theta)\right]^2},
\end{equation}
where $J_n(...)$ is the n-oder Bessel function. 
The oscillatory behavior of interlayer resistivity is directly seen from Eq. (\ref{LNL4}).
Indeed, at the commensurate directions of a magnetic field (\ref{commensurate}), where
$\omega_c(\theta,\varphi) = n \omega_b(\theta)$, there appear maxima of conductivity (\ref{LNL4}),
which lead to resistivity minima in 
$\rho_{zz}(H,\theta,\varphi)$ (see Fig. \ref{LNLFig}).
Note that Eq.(16) is an agreement with the previous results [47-50].

\subsection{LMA Effects}

The LMA phenomena are experimentally observed in a magnetic field, 
directed in $({\bf y},{\bf z})$ plane,
\begin{equation}
{\bf H}=(0,H\sin\alpha, H\cos\alpha),
\end{equation}
where $\alpha = 90^{\circ} - \theta$.
Therefore, Eq. (\ref{commensurate}) for the commensurate directions of a magnetic
field reduces to, 
\begin{equation}
\tan\alpha=n\left(\frac{b^*}{c^*}\right),
\end{equation}
where $n$ is integer.
In this case, where $\phi = 90^0$, Eq. (\ref{conductance}) for interlayer conductivity can be rewritten as,
\begin{eqnarray}
\label{LMA1}
\sigma_{zz}(H, \alpha)&\sim&\int\frac{{\rm d}y}{v_x(y)}\int_{-\infty}^0{\rm d}u
\frac{\cos\left[N(\alpha)u\right]}{\omega_b(y+u,\alpha)}\nonumber\\
&&\ \ \ \cdot\exp\left[-\int_u^0
\frac{{\rm d}v}{\tau\omega_b(y+v,\alpha)}\right],
\end{eqnarray}
where $\omega_b(y,\alpha)=eHv_x(y)b^*\cos\alpha/c$, 
$\omega_c(y,\alpha)=eHv_x(y)c^*\sin\alpha/c$, and 
$N(\alpha)=\omega_c(y,\alpha)/\omega_b(y,\alpha)$. 
Eq. (\ref{LMA1}) can be transformed to the following expression
by integrating by parts,
\begin{eqnarray}
\label{LMA2}
\sigma_{zz}(H, \alpha)&\sim&\int\frac{{\rm d}y}{v_x(y)}\Bigg[1+N(\alpha)\int_{-\infty}^0{\rm d}u
\sin\left[N(\alpha)u\right]\nonumber\\
&&\ \ \ \cdot\exp\left[-\int_u^0
\frac{{\rm d}v}{\tau\omega_b(y+v,\alpha)}\right]\Bigg].
\end{eqnarray}
If we introduce the following notations, 
\begin{eqnarray}
&&f(y)=\frac{v_F}{v_x(y)}-1,\nonumber\\
&&h_b(H,\alpha)=\frac{e}{c}Hv_Fb^*\tau\cos\alpha,\nonumber\\
&&h_c(H,\alpha)=\frac{e}{c}Hv_Fc^*\tau\sin\alpha,
\end{eqnarray}
where $1/v_F=\langle 1/v_x(y)\rangle_{p_y}$, 
then Eq. (\ref{LMA2}) becomes,
\begin{eqnarray}
\label{LMA3}
&&\sigma_{zz}(\theta)\sim\int{\rm d}y\left(1+f(y)\right)\nonumber\\
&&\cdot\Bigg[1+h_c(H,\alpha)\int_{-\infty}^0{\rm d}u
\sin\left[h_c(H,\alpha)u\right]\nonumber\\
&&\cdot\exp\left[-\int_u^0
{\rm d}v\left(1+f(y+h_b(H,\alpha)v)\right)\right]\Bigg].
\end{eqnarray}
Integrating by parts one more time and taking into account a 
periodicity of function $f(y)$, we obtain the following expression 
for conductivity,
\begin{eqnarray}
\label{MACon}
&&\frac{\sigma_{zz}(H,\theta)}{\sigma_{zz}(H=0)}=1-h_c^2(H,\alpha)\int_{-\infty}^0
{\rm d}u\exp(u)\\
&&\!\!\!\!\cdot\cos\left[h_c(H,\alpha)u\right]\int_0^{2\pi}\frac{{\rm d}y}
{2\pi}\exp\left\{-\int_u^0{\rm d}u_1f[y+h_b(H,\alpha)u_1]\right\}\nonumber.
\end{eqnarray}

\begin{figure}[t]
\includegraphics[width=.4\textwidth]{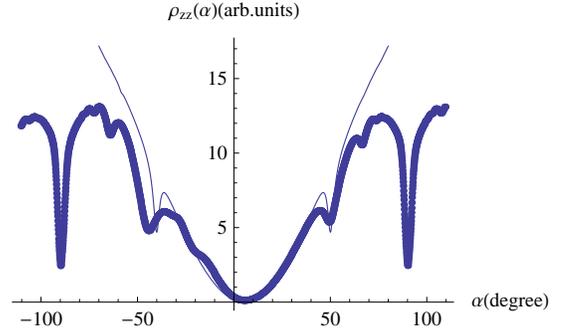}
\caption{\label{MAFig}Comparison of the theory (\ref{MACon1}) and experiment\cite{HaNaught} 
for the Lebed Magic Angle (LMA) phenomena in (TMTSF)$_2$ClO$_4$, 
where $g(p_yb^*)=[\cos^2(p_yb^*) +(\Delta/2t_b)^2]^{1/2}$ 
in Eq.(\ref{Spectrum}). 
Thin solid curve is numerically calculated from Eq. (\ref{MACon1}), with band 
parameters $t_a/t_b=8.5$ and $\Delta/2t_b=0.1$. 
For the calculations, we have used $\omega_b(0)\tau=15$ and $c^*=2b^*$.
 There exist good qualitative and quantitative agreement between the theory
 and experiment for $0^{\circ} < \alpha <  60^{\circ}$.
 For negative values of $\alpha$, agreement between the theory and experiment
 is worse, perhaps, due to some problems with the experimental measurements.}
\end{figure}

In the so-called clean limit, where $h_b(H,\alpha)\gg 1$, Eq. (\ref{MACon}) 
can be significantly simplified. 
Below, we introduce the 
Fourier transform of function $f(y)$,
\begin{equation}
f(y)=\sum_{n=1}^{+\infty}A_n\cos(ny).
\end{equation}
In the clean limit, the last exponential function in Eq. (\ref{MACon}), 
whose argument is inversely proportional to $h_b(H)$, 
can be expanded as,
\begin{eqnarray}
\label{MACon2}
	&&\!\!\!\!\!\exp\left\{-\int_u^0{\rm d}u_1f[y+h_b(H,\alpha)u_1]\right\}=1\nonumber\\
	&&-\int_u^0{\rm d}u_1\sum_{n=1}^{+\infty}A_n\cos\left[n(y+h_b(H,\alpha)u_1)\right]\nonumber\\
	&&+\frac{1}{2}\int_u^0{\rm d}u_1\int_u^0{\rm d}u_2\sum_{n,m=1}^{+\infty}
	A_nA_m\cos\left[n(y+h_b(H,\alpha)u_1)\right]\nonumber\\
	&&\cdot\cos\left[m(y+h_b(H,\alpha)u_2)\right],
\end{eqnarray}
where the higher order terms are discarded. 
After integration with respect to variable $y$, the second term in Eq.(\ref{MACon2}) vanishes,
whereas, in the third term, only contributions with $n=m$ retain.
Finally, the interlayer conductivity can be represented as, 
\begin{eqnarray}
\label{MACon1}
&&\!\!\!\!\!\frac{\sigma_{zz}(H,\alpha)}{\sigma_{zz}(H=0)}=\frac{1}{1+h_c^2(H,\alpha)}\nonumber\\
&&-\tan^2\alpha\left(\frac{c^*}{2b^*}\right)^2\sum_{n=1}^{\infty}\frac{A_n^2}{n^2}
\Bigg(\frac{2}{1+h_c^2(H,\alpha)}\nonumber\\
&&-\frac{1}{1+\left[h_c(H,\alpha)-nh_b(H,\alpha)\right]^2}\nonumber\\
&&-\frac{1}{1+\left[h_c(H,\alpha)+nh_b(H,\alpha)\right]^2}\Bigg).
\end{eqnarray}
Note that Eq.(26) is an agreement with our previous results [32,33].

In Fig. \ref{MAFig}, we compare the theory (\ref{MACon1}) with the experimental data\cite{HaNaught} for
the LMA phenomenon.
It is important that, for the calculations, we have used in Eq.(\ref{MACon1}) the same
values of the parameters as for the calculations of the LNL phenomenon
[see Eq.(\ref{LNL1}) and Fig. \ref{LNLFig}].
Fig. \ref{MAFig} demonstrates good qualitative and quantitative agreement between
the theory and experiment in the wide range of the angles: 
$0^{\circ} < \alpha < 60^{\circ}$.
For $|\alpha | > 60^{\circ}$, there appear significant deviations from the experimental
behavior$^{57}$.
One possible reason for that is a breakdown of Eq.(\ref{MACon1}) for large values of angle
$\alpha$, where the so-called clean limit approximation is not valid.
Another possible reason for the deviations is that, at high values of angle $\alpha$
(i.e., at high in-plane projections of a magnetic field), there may occur Fermi-liquid$^{59}$ 
or non Fermi-liquid$^{60}$ decoupling of the conducting layers, where
the Boltzmann kinetic equation is not valid any more.

\section{Conclusion}

In this work, we propose a unification theory for angular magnetoresistance 
oscillations, experimentally observed in Q1D organic conductors. 
We analytically solve the Boltzmann equation in the extended Brillouin zone
and find analytical formula, which describes interlayer resistivity.
We show that, in two important limiting cases, this formula reduces to the
expressions, previously obtained to describe the LNL and LMA phenomena
in resistivity.
Numerical results, obtained from these expressions, are shown to be in good
agreement with the experimental data, obtained on (TMTSF)$_2$ClO$_4$
conductor, in a broad rage of magnetic field directions.
On the other hand, a comparison of the theory with the LMA experimental 
data reveals significant discrepancy between the theory and experiment for
directions of a magnetic field close to the conducting layers.
We suggest that this discrepancy may be due to decoupling of the conducting
layers in a parallel magnetic field. 

\section{ACKNOWLEDGMENTS}

One of us (A.G.L.) is thankful to N.N. Bagmet (Lebed) for useful discussions.
Both of us are thankful to Heon-Ick Ha and M.J. Naughton for supplying us by 
experimental data on angular magnetic oscillations\cite{HaNaught} before their publication.
This work was supported by the NSF under Grant No. DMR-0705986.


\appendix

\section{Proof of Eq. (\ref{Solution2})}
\label{app}

From physical point of view, we have to find a particular solution of the
inhomogeneous partial differential equation (7), which is proportional to external 
electric field and periodic in the extended Brillouin zone with respect to both
variables, $y$ and $z$.
To find such solution, below we use the method of characteristics for the first
order partial differential Eq. (7).
Let us consider the following equation, which defines the characteristics:
\begin{equation}
\label{char}
\frac{{\rm d}z}{{\rm d}y}=\frac{\omega_c(y,\theta,\varphi)+\omega_c^*(\theta,\varphi)g^{\prime}(y)}
{\omega_b(y,\theta)},
\end{equation}
with initial condition being,
\begin{equation}
z(u)=z_0.
\end{equation}
Note that Eqs. (A1),(A2) have the following solution,
\begin{equation}
z(y)=z_0+\int_u^y\frac{{\rm d}v}{\omega_b(v,\theta)}
\left[\omega_c(v,\theta,\varphi)+\omega_c^*(\theta,\varphi)g^{\prime}(v)\right].
\end{equation}
At this point, Eq. (\ref{BE2}) can be considered as an ordinary differential equation,
\begin{equation}
eEv_z^0\sin z_0-\omega_b(y,\theta)\frac{{\rm d} \Psi(y,z_0)}{{\rm d}y}=
\frac{\Psi(y,z_0)}{\tau}.
\end{equation}
For our purpose, we choose the following solution of Eq. (A4),
\begin{equation}
\Psi(y,z_0)=eEv_z^0\int_{-\infty}^y\frac{{\rm d}v}{\omega_b(v,\theta)}\sin z_0
\exp\left\{-\int_v^y\frac{{\rm d}w}{\omega_b(w,\theta)\tau}\right\}.
\end{equation}
If we take into account that $z_0$ is given by Eq.(A3), we can obtain:
\begin{eqnarray} 
&&\Psi(y,z)= eEv_z^0\int_{-\infty}^y\frac{{\rm
d}u}{\omega_b(u,\theta)}\nonumber\\
&&\cdot\sin
\Bigg[z-\int_u^y\frac{{\rm
d}v}{\omega_b(v,\theta)}\Big[\omega_c(v,\theta,\varphi)\nonumber\\
&&+\omega_c^*(\theta,\varphi) g^{\prime}
\left(v\right)\Big]\Bigg]\exp\left[-\int_u^y\frac{{\rm d}v}{\tau\omega_b(v,\theta)}\right].
\end{eqnarray}
From Eq. (A6), it is directly seen that the obtained solution of Eq.(7) is
proportional to electric field and periodic with respect to variable $z$.
It is possible to make sure that it is also periodic with respect to variable $y$.
For this purpose, it is necessary to take into account that the functions 
$\omega_b(y, \theta)$, $\omega_c(y, \theta,\phi)$, and $g(y)$ are periodic
with respect to variable $y$ in the extended Brillouin zone.
After changes of integration variables, $u\rightarrow y+u$ and 
$v\rightarrow y+v$, we can finally obtain Eq. (9) from Eq. (A6).


$^*$ Also at: Landau Institute for Theoretical Physics, 2 Kosygina Street,
117334 Moscow, Russia.

\end{document}